
\input harvmac.tex
%
%
%
%
%
%
%
%
\def\nopage0{\pageno=0 \footline={\ifnum\pageno<1 {\hfil} \else
              {\hss\tenrm\folio\hss}   \fi}}
%
%

%
%
\def\sqr#1#2{{\vcenter{\hrule height.#2pt
      \hbox{\vrule width.#2pt height#1pt \kern#1pt
         \vrule width.#2pt}
      \hrule height.#2pt}}}

%
%
\def \d2dots{\mathinner{\mkern1mu\raise1pt\vbox{\kern7pt\hbox{.}}\mkern2mu
\raise4pt\hbox{.}\mkern2mu\raise7pt\hbox{.}\mkern1mu}}
%
%
\def\smdrct#1{{\vcenter{\hbox{\vrule width.4pt height#1pt}}\kern-1.5pt\times}}

%
%

%
%

                    \def\cU{{\cal U}}

%
%

\def\CC{\rlap {\raise 0.4ex \hbox{$\scriptscriptstyle |$}}
\hskip -0.1em C}
\def\FF{\hbox to 8.33887pt{\rm I\hskip-1.8pt F}}
\def\NN{\hbox to 9.3111pt{\rm I\hskip-1.8pt N}}
\def\PP{\hbox to 8.61664pt{\rm I\hskip-1.8pt P}}
\def\QQ{\rlap {\raise 0.4ex \hbox{$\scriptscriptstyle |$}}
{\hskip -0.1em Q}}
\def\RR{\hbox to 9.1722pt{\rm I\hskip-1.8pt R}}
\def\ZZ{\hbox to 8.2222pt{\rm Z\hskip-4pt \rm Z}} 
\def\demi{{1\over 2}}
%
%
%
\def\typeA{type ${\cal A}$}
\def\typeB{type ${\cal B}$}

\def\TB{{\cal B}\,}
\def\uq{\cU_q(sl(2))}
\def\Id{{\rm Id}}
%
%
%
%
\lref \rAkinetic { D. Arnaudon, {\sl Composition of kinetic momenta:
the ${\cal U}_q(sl(2))$ case,}
preprint CERN-TH.6730/92, to be published in Commun. Math. Phys.  }
\lref \rADHR {  F.C. Alcaraz, M. Droz, M. Henkel and V. Rittenberg,
{\sl Reaction-diffusion Processes, Critical Dynamics and Quantum Chains,}
Preprint UGVA-DPT 1992/12-799 Bonn-HE (1992).}
\lref \rBGS { A. B\'erkovich, C. G\'omez and G. Sierra,
{\sl Spin-anisotropy commensurable chains: quantum group symmetries
and $N=2$ SUSY,} Preprint Madrid (1993).}
\lref \rCou { M. Couture,
{\sl On some quantum $R$ matrices associated with representations of
$U_q(sl(2,\CC))$ when $q$ is a root of unity,}
J. Phys. A: Math. Gen. {\bf 24} (1991) L103, and {\sl Quantum
symmetries associated with the Perk--Schultz model,}
J. Phys. A: Math. Gen. {\bf 25} (1992) 1953.}
\lref \rGSii { C. G\'omez and G. Sierra,
{\sl New integrable deformations of higher spin Heisenberg-Ising chains,}
Phys. Lett. {\bf B285} (1992) 126.}
\lref \rHR { H. Hinrichsen and V. Rittenberg, {\sl A
two-parameter deformation of the $SU(1/1)$ superalgebra and the XY quantum
chain in a magnetic field,}
Phys. Lett. {\bf B275} (1992) 350.}
\lref \rInfAnalysis {
{\sl Infinite Analysis,}
Proceedings of the RIMS Research Project 1991, edited by A. Tsuchiya,
T. Eguchi and M. Jimbo (World Scientific).}
\lref \rLushnikov {  A.A. Lushnikov,
{\sl Binary reaction $1+1\rightarrow 0$ in one dimension,}
Sov. Phys. JETP {\bf 64} (1986) 811 and
Phys. Lett. {\bf 120A} (1987) 135.}
\lref \rNijs { M. den Nijs, {\sl The domain wall theory of
two-dimensional commensurate-incommensu\-rate phase transitions,}
in {\sl Phase
transitions and critical phenomena, Vol. 12,} C. Domb and J.L.
Lebowitz (eds.), Academic Press, New York (1988).}
\lref \rRA {  P. Roche and  D. Arnaudon,
{\sl Irreducible representations of the quantum
analogue of $SU(2)$,}
Lett. Math. Phys. {\bf 17} (1989) 295.}
\lref \rRosC { M. Rosso, {\sl Quantum groups at root of 1 and tangle
invariants,} in Topological and geometrical methods in field theory,
Turku, Finland May-June 1991, ed. by J. Mickelsson and O. Pekonen.}
\lref \rPerkSchultz {  J.H.H. Perk and C.L. Schultz,
{\sl New families of commuting transfer matrices in $q$-state vertex models,}
Phys. Lett. {\bf 84A} (1981) 407.}
\lref \rSalii { H. Saleur, {\sl Symmetries of the XX chain and applications,}
in Proceedings of {\sl Trieste conference on recent developments in
conformal field theories,} (1989). }
\lref \rSiggia { E. Siggia, {\sl Pseudospin formulation of kinetic
Ising models,} Phys. Rev. {\bf B16} (1977) 2319.}
%
%
\line{\hfil                                                   CERN-TH.6786/93}
\Title{}{\vbox{\centerline{Quantum Chains with $\uq$ Symmetry}
      \vskip2pt\centerline{and Unrestricted Representations}} }

\centerline{Daniel Arnaudon$^a$
\footnote{$^b$} {arnaudon@cernvm.cern.ch}
and Vladimir Rittenberg$^c$
\footnote{$^d$} {unp01@ibm.rhrz.uni-bonn.de}
}
\bigskip
{\it
\centerline{$^a$ Theory Division, CERN}
\centerline{1211 Gen\`eve 23, Switzerland}
\bigskip
\centerline{$^c$ Physikalisches Institut, Universit\"at Bonn}
\centerline{Nu{\ss}allee 12, D-5300 Bonn 1, Germany}
}
\bigskip
\bigskip
\centerline{\bf Abstract}
\bigskip
We consider two-state ($q^2=-1$) and three-state ($q^3=1$) one-dimensional
quantum spin chains with $\uq$ symmetry. Taking unrestricted (\typeB)
representations (periodic, semi-periodic and nilpotent), we show which
are the necessary conditions to obtain a Hermitian Hamiltonian.

\vfill

\leftline{CERN-TH.6786/93}
\leftline{January 1993}
\leftline{hep-th/9302088}

\baselineskip=17pt

\Date{}

One-dimensional quantum chains appear in equilibrium statistical
mechanics \refs{\rInfAnalysis, \rNijs}
and in non-equilibrium master equations which
describe annihilation-diffusion processes  \rLushnikov\ or critical
dynamics \rSiggia. A detailed description of the last two topics is
given in Ref. \rADHR. In problems of equilibrium, one is
often (but not always!) interested in Hermitian Hamiltonians, whereas in
non-equilibrium problems one is interested in Hamiltonians where all
the matrix elements are real since they are related to probabilities.
In the present note we examine the possibility of obtaining (up to
boundary terms) Hermitian or real Hamiltonians with two and three
states using $\uq$ unrestricted representations. The case of
restricted representations is already known \rPerkSchultz. Note that
at
this point we do not ask any questions about the integrability  of the
Hamiltonian.
\medskip
\medskip
\noindent {\bf   Definitions}
\medskip
The quantum algebra
$\uq$ is defined by the generators $k$, $k^{-1}$, $e$, $f$,
and the
relations
\eqn\eSL{
\eqalign{
& kk^{-1}=k^{-1}k=1 ,\cr
& [e,f]={k-k^{-1} \over q-q^{-1}}, \cr }
\qquad \qquad
\eqalign{
& kek^{-1}=q^2 e ,\cr
& kfk^{-1}=q^{-2}f .\cr}}

The
coalgebra structure is given by the coproduct
\eqn\eDelta {
\eqalign {
& \Delta(k)=k\otimes k \cr
& \Delta(e)=e\otimes 1      + k\otimes e \cr
& \Delta(f)=f\otimes k^{-1} + 1\otimes f \;,\cr  }}

\medskip
\noindent {\bf   Centre of $\uq$ when $q$ is a root of unity}
\medskip
Take $q$ such that $q^{m'}=1$ and define $m$ such that $m=m'/2$ if
$m'$  is even, $m=m'$ otherwise. Then the centre of $\uq$ is generated
by $e^m$, $f^m$, $k^{\pm m}$ and $C$ with
\eqn\eCas{C=fe+(q-q^{-1})^{-2}\left( qk+q ^{-1}k^{-1} \right)\;,}
altogether satisfying the polynomial relation
\eqn\ePmC{P_m(C)=e^m f^m + q^m {k^m +k^{-m} \over (q-q^{-1})^{2m}} \;,}
where
\eqn\ePm{P_m(X)=
{2\over (q-q^{-1})^{2m}} T_m\left( \demi (q-q^{-1})^{2} X \right)\;,}
$T_m$ being  the $m$-th Chebychev polynomial of
the first kind
\eqn\eCheb{T_m(X)= \cos(m \arccos X).}

\medskip
\noindent {\bf  Unrestricted, or \typeB\ representations of $\uq$}
\medskip
For the construction of our quantum spin chains, we will use the type
of irreducible representations (irreps)
that exists only when $q$ is a root of unity \rRA.
We call these irreps {\it \typeB\ } irreps (these representations all have
dimension $m$ and depend on continuous complex parameters), as
opposed to \typeA\ irreps corresponding to $q$-deformations of
ordinary representations of $\cU(sl(2))$ (which have dimension $\le
m$).

On these irreps, the central elements
$e^m$, $f^m$, $k^{\pm 1}$ and $C$ take the values
$x$, $y$, $z^{\pm 1}$, and $c$ lying on a 3-dimensional manifold
because of relation \ePmC.

$\lambda$ being an $m$-th root of $z$,
the \typeB\  representation denoted in the
following by $\TB(x,y,z,c)$
is given, in the basis $\{v_{0},...,v_{m-1}\}$, by
\eqn\eTypeB{
\cases {
k v_{p}=\lambda q^{-2p} v_{p} & for $0\le p \le m-1$ \cr
f v_{p}= \alpha_p v_{p+1} & for $0\le p \le m-2$ \cr
f v_{m-1}=\alpha_{m-1} v_0 \cr
e v_{p}=\alpha_{p-1}^{-1}
        \left( c - {1 \over (q-q^{-1})^{2}} \left( \lambda q^{-2p+1} +
        \lambda^{-1}  q^{2p-1} \right) \right) v_{p-1}
      & for $1\le p \le m-1$  \cr
e v_{0}= \alpha_{m-1}^{-1}
         \left( c - {1 \over (q-q^{-1})^{2}} \left( \lambda q +
         \lambda^{-1}  q^{-1} \right) \right)  v_{m-1}. \cr
}}
where $\prod_{p=0}^{m-1} \alpha_p =y$.

The freedom left by the choice of the $\alpha_p$ satisfying $\prod
\alpha_p=y$, corresponding to rescalings of the basis, will be used in
the following to explore more easily equivalent Hamiltonians for the
quantum chains.

We will distinguish three types of \typeB\ irreps:
\item{a)} { Periodic
representations, corresponding to injective action of $e$ and $f$,
i.e. to $xy\neq0$;}
\item{b)} { Semi-periodic representations, for which either $x$
or $y$ is $0$;}
\item{c)} { Nilpotent representations, corresponding to $x=y=0$
with only one complex parameter $z$.}

The nilpotent representations are also representations of the quantum
algebra generated by $e$, $f$ and $h$, where $k=q^h$. (Note that this
is not true in the other cases since $[h,e^m]=0$ on a representation
implies $e^m=0$ on it, and idem for $f^m$.) The logarithm of $k$ on
nilpotent representations is actually well defined, since the highest
weight and the lowest weight provide a cut in the values of $k$.

\medskip
\noindent {\bf  Quantum spin chains}
\medskip
We define a quantum spin chain with $\uq$ symmetry as follows: to each
site $j=1,...,L$ of the chain, we assign a \typeB\ representation
$\pi_j=\TB(x_j,y_j,z_j,c_j)$. We write the Hamiltonian
\eqn\eHam{H=\sum _{j=1}^{L-1} \Id\otimes ...\otimes \Id \otimes H_j \otimes ...
\otimes \Id }
with $H_j$ acting on sites $j$ and $j+1$ as
\eqn\eHj{H_j=(\pi_j\otimes\pi_{j+1})\left(Q_j(\Delta(C))\right)}
where $C$ is the quadratic Casimir \eCas\ and $Q_j$ is a polynomial of
degree $d< m$.

This Hamiltonian is by construction $\uq$ invariant.

\medskip
\noindent {\bf  On tensor products of \typeB\ irreps}
\medskip
On the tensor product of two \typeB\ irreps characterized by the
parameters $(x_1,y_1,z_1,c_1)$ and $(x_2,y_2,z_2,c_2)$, the central
elements
$\Delta(e^m)$, $\Delta(f^m)$ and $\Delta(k^{m})$ take
the (scalar) values
\eqn\exyz{
\eqalign{x&=x_1+z_1 x_2 ,       \cr
         y&=y_1 z_2^{-1}+y_2,   \cr
         z&=z_1 z_2.            \cr }}
The possible values for $\Delta(C)$ are given by the relation \ePmC.
More precisely, if we define $\zeta$ by
\eqn\eZeta{xy+q^m {z+z^{-1} \over (q-q^{-1})^{2m}} =
{\zeta ^m + \zeta ^{-m} \over (q-q^{-1})^{2m}}, }
then the possible values for $\Delta(C)$ are
\eqn\eCl{c_l={\zeta q^{2l} + \zeta^{-1} q^{-2l} \over (q-q^{-1})^{2}}
\qquad l=0,...,m-1.}
Then, if $\zeta^{2m}\neq 1$ we have the fusion rule
\eqn\eBBi{
\TB(x_1,y_1,z_1,c_1)\otimes \TB(x_2,y_2,z_2,c_2)  =
\bigoplus _{l=0}^{m-1}
\TB\left(x,y,z,c_l \right)
\;.}
With this, we know that the states of the whole chain lie in a sum of
$m$-dimensional multiplets, at least for generic values of the
parameters \rAkinetic.

\medskip
\noindent {\bf  Opposite coproduct and algebraic curve}
\medskip
We see from the fusion rules that the tensor product
$\pi_2\otimes \pi_1$
is equivalent to
$\pi_1\otimes \pi_2$
iff
\eqn\eCurve{
{x_1\over 1-z_1}={x_2\over 1-z_2}
\quad ,\qquad
{y_1\over 1-z_1^{-1}}={y_2\over 1-z_2^{-1}}\;.}
These conditions will be imposed on each pair of consecutive
representations of the chain. In other words, the parameters of all
the representations will lie on the same manifold defined by
\eqn\eCurveii{\eqalign{x&=A(1-z) \cr y&=B(1-z^{-1})\cr}}
for fixed $A,B$ (possibly $0$ or infinite. Note that $A$ or $B$
infinite means that $z$ is fixed to $1$ and that $x$ or $y$ are free.)

\medskip
\noindent {\bf  Some general results for the $m$-state quantum chains}
\medskip
\item {--} {A sufficient condition for the existence of a $U(1)$
invariance is obtained when $x_j=y_j=0$ for all $j$'s,
i.e. $A=B=0$, that is for
nilpotent representations.
As explained before, the generator $h$ is well-defined on each
representation, and the Hamiltonian commutes with
\eqn\eUone{\Delta^{(L)}(h)=\sum_{j=1}^L
\Id\otimes ...\otimes h\otimes ...\otimes \Id }
which generates a $U(1)$ symmetry.}
\medskip
\item{--} {A sufficient condition for having a real spectrum is the
following: if the parameters $x_j$, $y_j$, and $z_j$ are such that
\eqn\eReal{(x_j+z_jx_{j+1})(y_jz_{j+1}^{-1}+y_{j+1})+q^m
{z_jz_{j+1}+z_j^{-1}z_{j+1}^{-1} \over (q-q^{-1})^{2m}}
={\xi_j ^m + \xi_j ^{-m} \over (q-q^{-1})^{2m}}}
is a real number between $-2$ and $2$, then $\Delta(C)$ on sites
$j,j+1$ will have real eigenvalues (given by \eCl\ with
$\zeta=\xi_j$). If the polynomials $Q_j$ used to define the
Hamiltonian have real coefficients, then the spectrum is real.}

\medskip
\noindent {\bf  Two-state quantum chain: $q=i$, $m=2$}
\medskip
Let us consider a Hamiltonian constructed as above, with first finite
values for $A$ and $B$ in \eCurveii. Then  $H_j$ is
\eqn\eHamii{H_j=F_j\left( \rho_j \sigma_j^z + \rho_j^{-1} \sigma_{j+1}^z
   +\eta_j\sigma_j^x\sigma_{j+1}^x   +\eta_j^{-1}\sigma_j^y\sigma_{j+1}^y
\right)}
with $\eta_j=1$, $\rho_j=\left({z_{j+1}-1 \over 1-z_j^{-1}} \right)^{1/2}$
and $F_j$
is an arbitrary constant (note that the freedom in the choice of $Q_j$
corresponds to a redefinition of $F_j$). For the special choice
$\rho_j=\rho$ independent of $j$, we recover the $\cU_q(sl(1/1))$
supersymmetric Hamiltonian of Ref. \rSalii\ with $q=\rho$. It is
interesting to notice that the symmetries of this chain can now be
understood in a new way through \typeB\ representations of $\uq$.

{\it Remark}: The case of nilpotent representations is included here: it
corresponds to $A=B=0$. It has nothing remarkable since $A$ and $B$
do not change Eq. \eHamii.

\medskip
Suppose now that \eCurveii\ is satisfied with $A$ and $B$ infinite,
i.e. $z_j=1$, $\forall j$, and $x_j$, $y_j$ are free. Then we obtain
the Hamiltonian \eHamii\ with
$\rho_j=\left( {x_{j+1} y_{j+1} \over x_{j} y_{j} }\right) ^{1/4}$ and
$\eta_j=\left( {x_{j} y_{j+1} \over x_{j+1} y_{j} }\right) ^{1/4}$.
If we take $\rho_j=\rho$ and $\eta_j=\eta$ independent of $j$ we find
the Hamiltonian discussed in Ref. \rHR. This Hamiltonian is invariant
under a two-parameter ($\alpha$ and $\beta$ with $\alpha=\rho/\eta$
and $\beta=\rho\eta$) deformation of a superalgebra with two odd
generators. As for the $\cU_q(sl(1/1))$ case, here we understand the
symmetries in a different way.

{\it The crucial
feature here  is that the original parameters ${x_{j+1}\over x_j}$ and
${y_{j+1}\over y_j}$, which were parameters of representations (we had
$q=i$ in $SU(2)_q$ so no continuous deformation of the Lie algebra)
appear as deformation parameters of super Lie algebras. }
An analogous phenomenon was already observed in \refs{\rCou, \rRosC}.

On the whole chain, $\Delta^{(L)}(e)^2$, $\Delta^{(L)}(f)^2$ and
$\Delta^{(L)}(k)^2$ take the scalar values
$x_1\left( 1-\alpha^{2L}\over 1-\alpha^2\right)$,
$y_1\left( 1-\beta^{2L}\over 1-\beta^2\right)$ and $1$.
So for generic values of the parameters (of the representations)
$\alpha$ and $\beta$, the total representation will be periodic.
However, for $\alpha $ or $\beta$ a root of unity (say
$\alpha^{2L_0}=1$; $L_0$ having nothing to do with $m=2$!), the total
representation will lose its periodicity for chains of lengths $L$
which are multiples of
$L_0$,  and the total representation will contain indecomposable
subrepresentations.

\medskip

All the  Hamiltonians obtained with \typeB\ representations of $\uq$ with
$q=i$ are hence integrable \rHR, and are Hermitian if $\rho$ and
$\eta$ are real. If $\rho$ is on the unit circle, the Hamiltonian is
Hermitian up to boundary terms.

\medskip
\noindent {\bf  Three-state quantum chain: $q^3=1$, $m=3$}
\medskip
We consider
$H_j=a_2^{(j)} \Delta(C)^2 + a_1^{(j)} \Delta(C) + a_0^{(j)}$
with the freedom for the choice of $a_0^{(j)}$, $a_1^{(j)}$,
$a_2^{(j)}$ and for the parameters $(x_j,y_j,z_j,c_j)$ of the
representations $\pi_j$, as well as for the $\alpha_p^{(j)}$ (freedom
in the bases of the representations). We then look for the conditions
on these parameters for having a Hermitian Hamiltonian, or for having
a Hamiltonian with real matrix elements. A tedious calculation proves
that the following conditions are  necessary conditions for both cases:
\item{--} {All the representations are nilpotent, i.e. $x_j=y_j=0$
($\forall j$). The representations are then completely characterized
by their highest weights $\lambda_j$ (with $z_j=\lambda_j^3$ and
$c_j={q\lambda+q^{-1}\lambda^{-1}\over(q-q^{-1})^2}$).}
\item{--} {All the representations $\pi_j$ are equivalent to
$\pi_1\equiv \pi$, (i.e. $\lambda_j=\lambda$, $\forall j$).}
\item{--} {$|\lambda| =1$}

As explained before, the Hamiltonian will have a $U(1)$
symmetry since we have to consider nilpotent representations only.

A necessary and sufficient condition for the Hamiltonian to be
Hermitian
(up to boundary terms)
is
\eqn\ePhase{{\rm Re}\left( (q\lambda)^2 \right) \ge -{1\over 2}\;.}
The Hamiltonian density  $H_j$ is then written
\eqn\eHlambdai{
\eqalign{ H_j=
  D &\left( E^{01}\otimes E^{10} + E^{10}\otimes E^{01} \right)
+ H  \left( E^{12}\otimes E^{21} + E^{21}\otimes E^{12} \right) \cr
-   &\left( E^{02}\otimes E^{20} + E^{20}\otimes E^{02} \right) \cr
+ F &\left( E^{01}\otimes E^{21} + E^{21}\otimes E^{01}
          + E^{10}\otimes E^{12} + E^{12}\otimes E^{10} \right) \cr
}}
with
\eqn\eDFH{D={\lambda q^{-1}-\lambda^{-1}q \over q-q^{-1}} \qquad
          H=-{\lambda -\lambda^{-1} \over q-q^{-1}} \qquad
 F= {i  \over q-q^{-1}}\left(1+\lambda^2q^2+\lambda^{-2}q^{-2}\right)^{1/2}
}
for the non-diagonal part,
\eqn\eHlambdaii{A\otimes 1 +1\otimes A \qquad
\hbox{with} \qquad
A={1\over 2} {\rm Diag}\left(
    {\lambda^2 q^{-1}-\lambda^{-2}q \over q-q^{-1}} +1,2,
   -{\lambda^2 q^{-1}-\lambda^{-2}q \over q-q^{-1}} +1  \right)}
for the real diagonal part and
\eqn\Hlambdaiii{B\otimes 1 -1\otimes B \qquad
\hbox{with} \qquad
B={1\over 2} {(\lambda-\lambda^{-1})(\lambda q^{-1}-\lambda^{-1}q)
     \over q-q^{-1}}
   {\rm Diag}\left( 1 ,0, - 1\right)}
for the imaginary diagonal part (boundary term).
This Hamiltonian (up to the imaginary boundary term)
coincides with the one discovered by Gomez and Sierra \rGSii.
\medskip

The limit $\lambda \rightarrow 1$ of this Hamiltonian is
\eqn\eHun{
\eqalign{H_j=
 -&\left( E^{01}\otimes E^{10} + E^{10}\otimes E^{01} \right)
 - \left( E^{02}\otimes E^{20} + E^{20}\otimes E^{02} \right) \cr
 +&\left( E^{00}\otimes E^{11} + E^{00}\otimes E^{22}
         +E^{11}\otimes E^{00} + E^{22}\otimes E^{00} \right) \cr
 +&2\left( E^{11}\otimes E^{11} + E^{11}\otimes E^{22}
         +E^{22}\otimes E^{11} + E^{22}\otimes E^{22} \right) \cr
}}

This last Hamiltonian, obtained with $\lambda=1$,
is actually the only Hamiltonian (up to
equivalence transformations) which has only real matrix elements.
This special case is discussed in detail in \rADHR, where one
considers a two-species model of diffusion-annihilation processes.

\bigskip
When this work was completed, we received Ref. \rBGS, where the
q-chains based on nilpotent representations are studied.

\bigskip
{\bf Acknowledgements:} We would like to thank C. Gomez for many
discussions concerning Ref. \rGSii\ and for expressing his scepticism
about the existence of Hermitian Hamiltonians built with periodic
representations. He was proved right. We would also like to thank O.
Babelon and P. Roche for fruitful discussions, and T. Baker and M.
Scheunert for carefully reading the manuscript.

\listrefs
\bye